\documentclass[aps,pra,reprint,groupedaddress,longbibliography]{revtex4-1}
\usepackage{listings,amsmath,ams fonts,amssymb,graphicx,color,mathrsfs,cancel,braket,hyperref}

\newcommand{\e}{\textrm{e}^}
\newcommand{\E}{\mathcal{E}}

\newcommand{\norm}[1]{\left|\left| #1 \right|\right|}
\newcommand{\one}{\mathbb{I}}
\begin{document}

\title{Optimal Trotterization in universal quantum simulators under faulty control}

\author{George~C.~Knee}
\email[]{gk@physics.org}
\author{William~J.~Munro}

\affiliation{NTT Basic Research Laboratories, NTT Corporation, 3-1 Morinosato-Wakamiya, Atsugi, Kanagawa 243-0198, Japan}

\date{\today}

\begin{abstract}
Universal quantum simulation may provide insights into those many-body systems that cannot be described classically, and that cannot be efficiently simulated with current technology. The Trotter formula, which decomposes a desired unitary time evolution of the simulator into a stroboscopic sequence of repeated elementary evolutions, is a key algorithmic component which makes quantum simulation of dynamics tractable. The Trotter number $n$ sets the timescale on which a computer running this algorithm is switched from one elementary evolution to another. In the ideal case, the precision of the simulation can be arbitrarily controlled by increasing $n$. We study a more realistic scenario where each gate is applied imperfectly. The resultant tradeoff in errors leads to an ultimate limit on the precision of the simulation. We calculate the optimum Trotter number $n^*$ that achieves this limit, which is the minimum statistical distance from the actual simulation to the ideal one. 
\end{abstract}
\maketitle

\section{Introduction}
Simulation involves the control and study of a first physical system (the simulator) in order to provide insight into a second (the simuland). It is an indispensable tool in modern science. Under a classical computing paradigm, a huge variety of simulands are investigated using silicon based digital machines, which represent and process classical information stored as binary data in memory. For a broad class of problems, the simulation is~\emph{efficient} -- the simulator requires spatial and temporal resources scaling polynomially rather than exponentially in the size of the simuland. When the simuland is such that a fully quantum mechanical model is necessary to describe it, however, the memory requirement for a direct and exact simulation scales exponentially in the number of particles. Examples include quantum chemistry simulations~\cite{KassalWhitfieldPerdomo-Or2011} and simulations of condensed matter systems~\cite{BarendsLamataKelly2015,LanyonHempelNigg2011,MaDakicNaylor2011}. Often a mean-field approximation or cutoff in long-range entanglement can reduce the complexity and make the problem tractable once more. 

The idea of~\emph{quantum} simulation, on the other hand, where the computer can store and process~\emph{quantum} information, offers the potential to explore quantum simulands without making such assumptions. Quantum simulators are typically thought of as being analog or digital~\cite{BulutaNori2009}. Analog quantum simulators are purpose built, and have limited reprogrammability. They often process information stored in continuous variables, and therefore do not benefit from error correction. The concept is similar to the idea of an orrery, or a scale model. 

Digital or universal quantum simulators, however, are anticipated to operate under the quantum computation paradigm~\cite{Lloyd1996,Feynman1982,BrownMunroKendon2010,LaddJelezkoLaflamme2010}. Although they may in fact have more modest requirements, for our purposes they will be thought of as fully fledged quantum computers. The advantage of this type of simulation is that a single computer can be reprogrammed to perform any simulation, as long as it has enough memory. Perhaps more importantly, the error in the outcome of the calculation can be controlled by error-correction methods.

Quantum computers are typically constructed with two-level systems known as `quantum bits'. Such elements may be realised with trapped ions~\cite{LanyonHempelNigg2011}, superconducting circuits~\cite{HerasMezzacapoLamata2014,BarendsLamataKelly2015}, or other candidate systems~\cite{GeorgescuAshhabNori2014}. The formal requirements for quantum computation~\cite{DiVincenzo2000} dictate that a fiducial initial state can be prepared, that a universal set~\cite{BarencoBennettCleve1995,DeutschBarencoEkert1995,Lloyd1995} of operations can be applied, and that projective measurement is possible. 

Universal quantum simulation, being an instance of quantum computation, is therefore composed of preparation, evolution and readout stages.  Analyses concerning state preparation and data extraction can be found elsewhere~\cite{Zuniga-HansenChiByrd2012,PeruzzoMcCleanShadbolt2014}. {\color{black}The aim of this paper is to quantify the accuracy of the evolution stage when the control operations are faulty. To that end, in Section~\ref{simulation_algorithm} we discuss the algorithms used for quantum simulation and recap arguments concerning accuracy with ideal operations; in Section~\ref{noise_types} we discuss a general framework for modeling faulty operations; in Section~\ref{statistical_distance} we discuss various ways to quantify the accuracy of a quantum computation. Our main results are found in Section~\ref{results}, where we argue that a tradeoff between Trotter error and gate errors is generic to universal quantum simulation and exhibit this with several examples.} Our results show how to operate a faulty simulation in the optimal way, and the level of precision expected at this optimum.
\section{Simulation Algorithm}
\label{simulation_algorithm}
{\color{black}Even when a quantum computer with a universal gate set is available, one does not generally know how to combine the gates \emph{efficiently} to achieve a particular desired evolution.}

\subsection{Lloyd's algorithm}
Lloyd's algorithm~\cite{Lloyd1996} is a general but approximate solution to this problem, when the desired evolution is known to be generated by a local Hamiltonian: \begin{equation}
H=\sum_{j=1}^k H_j,
\end{equation}
where each of the $k$ component Hamiltonians $H_j$ has dimension less than some maximum (call this $g$). The true evolution is then approximated through a truncation of the Trotter~\cite{Trotter1959} formula:
\begin{equation}
U = \textrm{exp}[iHt]=\left(\prod_{j=1}^k \textrm{exp}[iH_j t/n]\right)^n+\ldots,
\label{trotter}
\end{equation}
where we set $\hbar=1$, and take $H$ to be time independent (although this can be straightforwardly relaxed). Further, we rescale our units such that time $t$ is dimensionless.

The advantage of Lloyd's approach is that the number of operations required is bounded by a number proportional to $t^2kg^2/\epsilon$, where $t$ is the simulation time, $g$ is the maximum dimension of the local Hamiltonians, and $\epsilon$ is the desired error~\cite{Lloyd1996}.  The total number $k$ of component Hamiltonians has a better-than-polynomial (rather than exponential) dependence on the particle number, making Lloyd's method `efficient'.

It is clear that for any finite value of $n$ (the `Trotter number'), the higher order terms in the above equation will be non-zero. Their neglect then leads to an error in the simulation - the simulator is not driven to the desired final state but to one that is nearby. As we shall see, `nearby' can be given a concrete mathematical and operational definition. 

Higher Trotter numbers result in trajectories of the simulator that result in a final state that is closer to the ideal. Increasing $n$ will generally increase the computational complexity -- more control operations. More time may not be required, however, because the gates are correspondingly shorter: although this appears not to be the case in error corrected implementations`\cite{BrownClarkChuang2006}. Others have considered the dependence of the number of gates and simulation time on the desired error~\cite{BrownClarkChuang2006}, but here our only concern is the overall accuracy of the simulation. Higher order approximants are available~\cite{JankeSauer1992,Suzuki1992}, or other techniques that exploit sparsity~\cite{BerryAhokasCleve2007}. The common attributes of these approaches are that they reduce the number of operations to polynomial in the particle number, and that their accuracy is controlled (improved) by increasing the number of applied operations. The number of Trotter steps necessary for quantum chemistry simulation was considered in Refs.~\cite{WeckerBauerClark2014,PoulinHastingsWecker2015}.

Throughout this paper we take $V_j=\text{exp}[iH_jt/n]$ as primitive operations applied to the simulator, although ultimately these primitives should be understood as being composed from gates drawn from the particular universal set that is available.

\subsection{Time-energy freedom}
Because of the $H\rightarrow aH, t\rightarrow t/a$ symmetry of the Schr\"odinger equation  $i \frac{\partial}{\partial t}|\psi\rangle=-H|\psi\rangle$, when simulating closed-system dynamics one benefits from the freedom to define 
\begin{align}
H_{\text{simulator}}&=aH_{\text{simuland}} \nonumber \\
t_{\text{simulator}}&=\frac{1}{a}t_{\text{simuland}},
\label{tef}
\end{align}
which will preserve the correspondence between the time evolutions of the simulator and simuland. The simulation time, therefore, may be chosen to be any duration as long as the appropriate global scaling of the energy of the control fields is also performed. Even classical simulators are rarely operated at a speed commensurate with their respective simulands-- weather patterns of several weeks are simulated in a matter of hours, and supercomputers spend months calculating chemical reaction dynamics over timescales many orders of magnitude smaller. In fact, because each unitary $\text{exp}[iH_jt/n]$ is decomposed into the natural elementary gate set of the simulator, one generally expects $a\neq1$. Clearly this freedom enables the computation to be sped up (or slowed down) by a constant factor~\footnote{A simulator cannot be operated in sublinear time: that is, it cannot take less than a time proportional to $t_{\text{simuland}}$~\cite{BerryAhokasCleve2007}.}, but more importantly we imagine that such freedom may prove very useful for reducing noise in implementations of quantum simulation, depending on the particular noise which dominates -- see below. It is worth noting that assuming this freedom is {\color{black} asking} more than is necessary for universal quantum computation -- nevertheless in many quantum computers we expect there to be at least a limited ability to perform the primitive operations at different physical speeds. 

\section{Noise types}
\label{noise_types}
The error in approximating the true evolution with the first term on the right-hand-side of $\eqref{trotter}$ is known to decay at worst as $\propto n^{-1}$~\cite{ReedSimon1980}, and as Lloyd put it `$n$ can always be picked sufficiently large to ensure the simulator tracks the correct time evolution to within any [desired nonzero accuracy]'~\cite{Lloyd1996}. The perfect control of any system (quantum or classical) is only ever an idealisation however. {\color{black}Little is known about the real-world situation -- although recently the Trotter decomposition has been shown to be stable in the sense that the overall error can be reduced to a fixed desired level if the precision of the individual steps is good enough~\cite{DhandSanders2014}}. Ref.~\cite{DurBremnerBriegel2008} is a study of the influence of noise on certain quantum simulations, calculating the average fidelity of the final state of the computer with the ideal. Here we derive analytical results for arbitrary (generally non-unitary and non-commuting) component Hamiltonians. Further, as we shall show, our results allow us to predict the ultimate performance of a faulty quantum simulator, and allow us to prescribe the optimum Trotter number to employ. Our results hold for a whole class of statistical distance metrics, including the most interesting~\emph{worst-case} metrics.

We construct `faulty Trotterized quantum channels', and consider the following generalised noise map, with $\mathcal{V}_j(\rho)=e^{iH_jt/n}\rho e^{-iH_jt/n}$ representing the component unitary processes of Trotterization (here describing the transformation of a $d\times d$ density matrix $\rho$ describing the quantum state of the simulator):
\begin{align}
\mathcal{E}^{\text{faultyTrotter}}(\rho)&=\bigcirc_{i=1}^n \bigcirc_{j=1}^k   \mathcal{E}_{ij} \circ \mathcal{V}_j(\rho)\nonumber\\
&=\bigcirc_{i=1}^n   \left(\ldots\circ\mathcal{E}_{i2} \circ \mathcal{V}_2 \circ \mathcal{E}_{i1} \circ \mathcal{V}_1\right)(\rho) \nonumber \\
& = \bigcirc_{i=1}^n \mathcal{E}^{\text{fTss}}_i (\rho).
\end{align}
The $\circ$ symbol is used here to denote the concatenation of quantum channels: here we have a triple concatenation (first due to faulty evolutions $\mathcal{E}_{ij}$ following the clean ones; second due to applying each of the $k$ local Hamiltonians in turn, and third by repeating this process $n$ times). The interleaved operations $\mathcal{E}_{ij}$ are unwanted evolutions; they carry an index $i$ to emphasise that they may vary over the course of the experiment, and there may also be a dependence on $\{H_j\}$, $t$ and $n$. In the course of our derivations it is useful to consider the map over a single Trotter iteration (faulty Trotter single step) $\mathcal{E}^{\text{fTss}}_i$.

 It will be convenient to consider the supermatrix representation of these maps: this is defined by $\mathbf{T}_{\mathcal{E}}\vec{\rho}=\overrightarrow{\mathcal{E}(\rho)}$, where $\vec{\rho}$ is the vectorized (column-stacked) density matrix and $\mathbf{T}_{\mathcal{E}}$ is a supermatrix of dimension $d^2$. This representation makes the semigroup structure of quantum maps apparent. The concatenation of channels is merely matrix multiplication: 
\begin{align}
\mathbf{T}^{\text{faultyTrotter}}&=\prod_{i=1}^n \prod_{j=1}^k   \mathbf{T}_{\mathcal{E}_{ij}}  \mathbf{T}_{\mathcal{V}_j}\nonumber\\
&=\prod_{i=1}^n   \left(\ldots\mathbf{T}_{\mathcal{E}_{i2}}  \mathbf{T}_{\mathcal{V}_2}\mathbf{T}_{\mathcal{E}_{i1}}  \mathbf{T}_{\mathcal{V}_1}\right) \nonumber \\
& = \prod_{i=1}^n \mathbf{T}^{\text{fTss}}_i .
\end{align}

We write the perfect implementation of the Trotter technique as $\mathcal{V}$ with unitary matrix $V=(\sqrt[n]{V})^n$ and supermatrix $\mathbf{T}_{\mathcal{V}}=(\sqrt[n]{\mathbf{T}_{\mathcal{V}}})^n$ -- essentially by removing all the faulty maps from the faulty Trotter channel.

We write the supermatrix representation of the ideal map $\mathcal{U}$ as $\mathbf{U}=U\otimes U^*$, and break up this unitary evolution as 
\begin{align}
\mathcal{U}(\rho) = \bigcirc_{}^n \sqrt[n]{\mathcal{U}}(\rho)
\end{align}
with the notation justified by the supermatrix describing each ideal Trotter step $\mathbf{U}=(\sqrt[n]{\mathbf{U}})^n$.

\section{Statistical distance measures for quantum channels}
\label{statistical_distance}
\begin{figure}
\includegraphics[width=8cm]{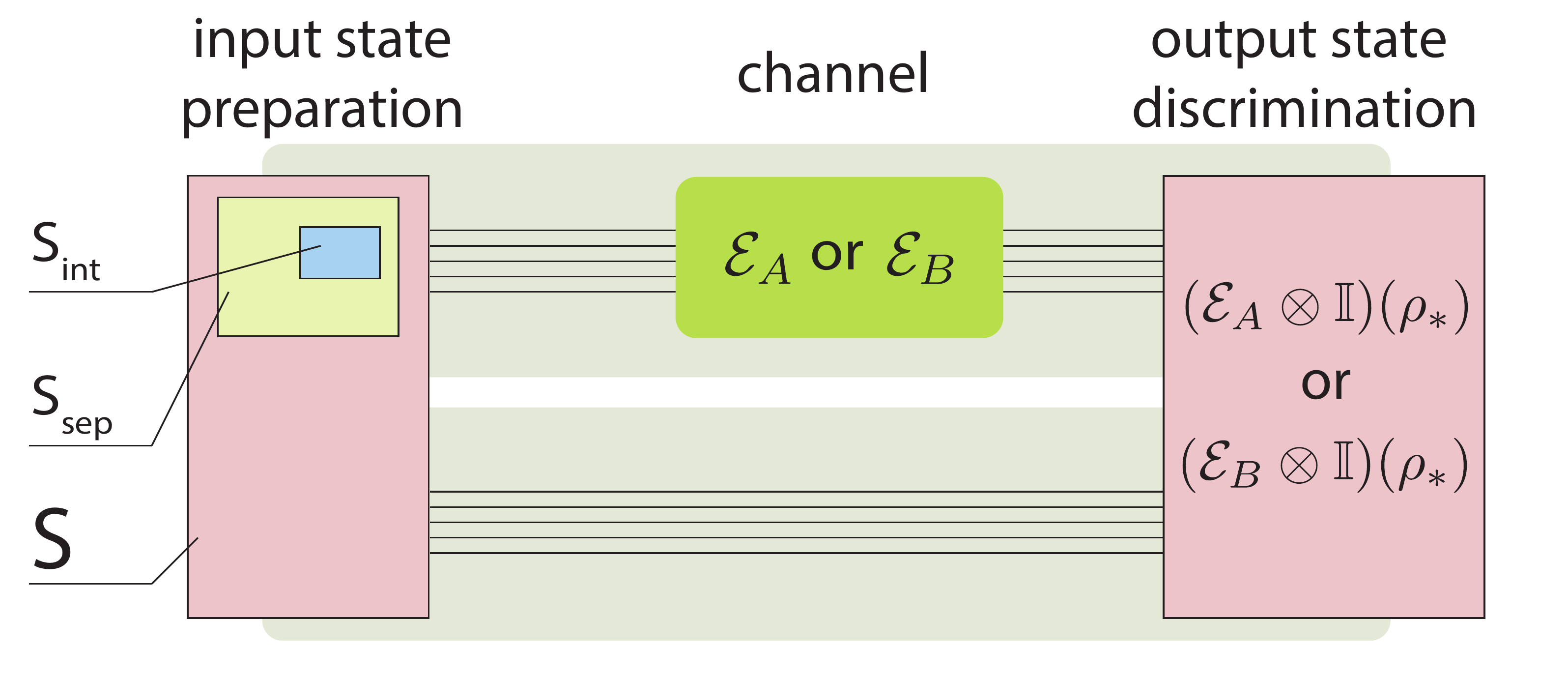}
\caption{(Color online) The distinguishability of two quantum channels $\mathcal{E}_A$ and $\mathcal{E}_B$ can be defined as the distinguishability of the outputs of each channel when the input is an optimally chosen quantum state $\rho_*$. The optimum is generally drawn from a set $S$ describing joint states of an enlarged system. The subset $S_\text{sep}$, contains states with no system-ancilla entanglement and a further subset $S_\text{int}$, which contains only physical states of interest. }
\label{sets}
\end{figure}

Hauke \emph{et al.} pose the question -- can we trust quantum simulators?~\cite{HaukeCucchiettiTagliacozzo2012}. Rather than give a binary answer, one can instead ask the question `to what extent can we trust quantum simulators'?  We aim to provide an answer to the above question in the form of a discrimination probability. To measure the faithfulness of quantum operations, one can appeal to generalisations of either classical fidelity or classical statistical distance between probability distributions. We prefer the latter here:
the \emph{trace distance} between quantum \emph{states} (density matrices) has an operational meaning because it determines the success probability for the state discrimination problem~\cite{Fuchs1996}. One is provided with either $\rho_A$ or $\rho_B$ with equal probability and is asked to guess which after a single measurement. The probability of success is
\begin{align}
p_{\text{distinguish}}=\frac{1}{2}+\frac{1}{4}||\rho_A-\rho_B||.
\end{align}
Here 
\begin{align}
||\rho_A-\rho_B||&=\text{Tr}|\rho_A-\rho_B|\nonumber\\
&=\text{Tr}\sqrt{(\rho_A-\rho_B)(\rho_A-\rho_B)^\dagger}.
\end{align}
The norm of the difference between the operators is a metric, giving the the quantum statistical distance between the operators. This definition implies that the discrimination is informed by measurement results arising from the optimum choice of {\color{black}POVM (Positive Operator Valued Measure), or generalised measurement procedure~\cite{NielsenChuang2004}}. In fact 
\begin{align}
||\rho_A-\rho_B||=2\max_{E}\text{Tr}(E(\rho_A-\rho_B)),
\end{align}
where $E$ is a positive operator (or POVM element).
A similar maximisation over states in turn induces a distance on quantum~\emph{channels}, defined as 
\begin{align}
||\mathcal{E}_A-\mathcal{E}_B||_{\diamond,1} &= \max _{\rho\in S,S_{\text{sep}}}||(\mathcal{E}_A\otimes \mathbb{I})(\rho)-
(\mathcal{E}_B\otimes \mathbb{I})(\rho)||.
\end{align}
The associated task is to submit an optimal initial state $\rho_*$ to undergo evolution under $\mathcal{E}_A$ or $\mathcal{E}_B$ (chosen at random), and to perform the state discrimination task (above) on the output state. For full generality the enlarged search space $S$ (having dimensions $d^2$) is needed to allow for the possibility of entanglement with an ancilla (of dimension no larger than that defined by the channels themselves) assisting in the channel discrimination task~\cite{KitaevShenVyalyi2002}. When the full search space $S$ is available, $||\cdot||_\diamond$ is the diamond norm~\cite{KitaevShenVyalyi2002}; otherwise, when the maximization is over states which factorize into system, ancilla states, the norm is the unstabilized induced trace-norm $||\cdot||_1$ (and the ancilla plays no role). A further restriction considered by Lloyd~\cite{Lloyd1996} would be to define the set of states of `interest' $S_{\text{int}}$: but this has the drawback of requiring detailed knowledge of the particular simulation at hand, and makes additional assumptions on the set of initial states of interest: see Figure~\ref{sets}.

These metrics then give the bias away from a half in the probability of discriminating between the real and ideal channels, given a single optimal initial state preparation (for the diamond norm over a larger space), a single channel use and a single sample from an optimally chosen measurement basis (again for the diamond norm on a larger space):
\begin{align}
p_{\text{distinguish}}\leq\frac{1}{2}+\frac{1}{4}||\mathcal{E}_A-\mathcal{E}_B||_{\diamond,1}.
\end{align}
They are therefore worst-case metrics. Yet another norm to consider is the $J$-norm $||\cdot||_J$: this provides a bound on the average trace distance over a uniformly measured state space~\cite{GilchristLangfordNielsen2005}. The $J$-distance between two quantum maps is merely the trace distance of the associated states in the Jamiolkowski isomorphism~\cite{Jamiolkowski1972}:
\begin{align}
\norm{\mathcal{E}_A-\mathcal{E}_B}_J = \norm{J(\mathcal{E}_A)-J(\mathcal{\mathcal{E}_B})},
\end{align}
where $J(\mathcal{E}) \propto \sum_{ij} \mathcal{E}(|i\rangle\langle j|) \otimes  |i\rangle\langle j | $ with the constant of proportionality such that $J(\mathcal{E})$ has unit trace. The $J$-distance is related to the average probability of discriminating the real and ideal simulations, given an optimal measurement:
\begin{align}
\bar{p}_{\text{distinguish}}\leq \frac{1}{2}+\frac{1}{4}\norm{\mathcal{E}_A-\mathcal{E}_B}_J.
\end{align}
For a comprehensive survey of available metrics, see Ref.~\cite{GilchristLangfordNielsen2005}~\footnote{In \cite{GilchristLangfordNielsen2005} they refer to the diamond norm as the $S$-distance.}. 

There are a number of senses in which these metrics are pessimistic measures. Firstly, for the worst-case metrics, an optimal input state (possibly entangled with an ancilla) may not arise in a quantum simulator: readout methods such as the phase estimation algorithm~\cite{Aspuru-GuzDutoiLove2005}, for example, do not employ such states. Currently, the full set of preparations and measurements at the disposal of a quantum computer are not made use of in quantum simulation algorithms: in the future, more involved procedures (including readout using an ancilla) may turn out to be more powerful at extracting information. Secondly, the maximisation over POVMs: in reality, the measurements are likely to be restricted to those observables necessary for calculating macroscopic quantities, e.g. the ground state energy of the system, the Shannon entropy, effective temperature, net magnetisation and so on. Consider that these measurement choices might be completely impervious to certain types of error, but yet the errors are ruthlessly sought out and exposed under the trace-norm or diamond-norm distances. In other words, there may be an equivalence class of (possibly mutually orthogonal) microstates with the same macrostate (or property of interest). An `incorrect' microstate could well be good enough for the purposes of the simulation (even yielding precisely the correct `answer') whilst being highly (or even perfectly) distinguishable from the correct state. Consider for example a very large simulator suffering only a bit flip on a qubit corresponding to the least significant digit of the readout register (minimal effect on readout result), or even a simulator which suffers a severe scrambling of phase relations prior to measurement in the computational basis (potentially zero effect on readout result). The game of simulation is not necessarily an adversarial one -- nevertheless due to the often random and potentially significant effect of errors, worst-case metrics are usually regarded as the most relevant measures to employ. 

All metrics satisfy some important properties which we will make use of: the triangle inequality $\norm{\mathcal{E}_A-\mathcal{E}_C}\leq\norm{\mathcal{E}_A-\mathcal{E}_B}+\norm{\mathcal{E}_B-\mathcal{E}_C}$ and convexity $\norm{\sum_ip_i\mathcal{E}_{i}}\leq \sum_ip_i\norm{\mathcal{E}_{i}}$. Another important property satisfied by all metrics considered in this paper is chaining: $\norm{\mathcal{E}_{A_1}\circ \mathcal{E}_{A_2} - \mathcal{E}_{B_1}\circ \mathcal{E}_{B_2}} \leq \norm{\mathcal{E}_{A_1}-\mathcal{E}_{B_1}}+\norm{\mathcal{E}_{A_2}-\mathcal{E}_{B_2}}$. The diamond norm and J distance are special in that they are \emph{stable}: they satisfy $\norm{\one\otimes\mathcal{E}}_{\diamond,J}=\norm{\mathcal{E}}_{\diamond,J}$. See Gilchrist \emph{et al.}~\cite{GilchristLangfordNielsen2005} for further details.

The convexity property is an important one, and is enough to ensure that the maximum statistical distance is achieved on a pure state~\cite{GilchristLangfordNielsen2005}. Further, although often understood as the inability to increase distinguishability through averaging (i.e. as a handicap in information processing tasks such as parameter estimation~\cite{NielsenChuang2004}), here it implies that the performance of a faulty quantum computation may be~\emph{improved} by repeating the computation and averaging the results. As we shall see below, this can lead to quite significant improvements, because the fluctuations may be suppressed.

 A related fact concerns appropriate worst-case benchmarks.  Whilst the ideal evolution $\mathcal{U}$ is unitary and will take a pure state to a pure state, a completely noisy channel $\mathcal{E}^{\rho\rightarrow\mathbb{I}/d}(\rho)=\mathbb{I}/d$ destroys the purity of any state (having dimension $d$). One can prove~\footnote{Due to the property of unitary invariance we can assume here that the unitary channel is the identity channel} that under the unstabilised trace norm one has 
\begin{align}
\norm{\mathcal{E}^{\rho\rightarrow\mathbb{I}/d}-\mathcal{U}}_{1}=2-\frac{2}{d},
\end{align}
whereas with the diamond norm~\cite{Sacchi2005} and $J$-distance (which we concentrate on for the remainder of this article) one has
\begin{align}
\norm{\mathcal{E}^{\rho\rightarrow\mathbb{I}/d}-\mathcal{U}}_{\diamond,J}=2-\frac{2}{d^2}.
\end{align}
These values represent benchmarks -- the distinguishability between the output of the ideal simulation and complete noise. Any simulation giving a higher distance that this will be worse than a completely random output! 
Note that both norms converge to the algebraic maximum of $2$ as $d\rightarrow\infty$.  The algebraic maximum is achieved, for example, by an unwanted bit flip.

\section{Results}
\label{results}
We are now in a position to calculate the accuracy of a realistic quantum simulator. To that end, we apply the norms introduced above to calculate the statistical distance from the ideal map to the faulty Trotter map. To see why we expect an optimum Trotter number to emerge in realistic universal quantum simulators, consider first a simple application of the triangle inequality to the distance between ideal and realistic channels over a single Trotter step:
\begin{align}
\norm{\sqrt[n]{\mathcal{U}}-\mathcal{E}^{\text{fTss}}} \leq \norm{\sqrt[n]{\mathcal{U}}-\sqrt[n]{\mathcal{V}}} + \norm{\sqrt[n]{\mathcal{V}}-\mathcal{E}^{\text{fTss}}}.
\end{align}

Next, assume the first term is bounded by a quantity $\propto n^{-s}$ for some $s\geq 2$: this captures the first order Trotter formula that we study here ($s=2$), as well as higher order expansions~\cite{JankeSauer1992}. Now assume the second term is a constant with respect to $n$. This immediately leads, via the chaining property, to
\begin{align}
D := \norm{\mathcal{U}-\mathcal{E}^{\text{faultyTrotter}}} \leq \frac{\mathscr{C}}{n^{s-1}} + \mathscr{D}n
\end{align}
for some constants $\mathscr{C}$ and $\mathscr{D}$. It is simple to show that the optimum Trotter number is
\begin{align}
\label{mainresult1}
n^* = \sqrt[s]{\frac{\mathscr{C}(s-1)}{\mathscr{D}}}
\end{align}
and the statistical distance at this optimum is
\begin{align}
\label{mainresult2}
D(n^*) = \frac{\mathscr{D}s}{s-1}\sqrt[s]{\frac{\mathscr{C}(s-1)}{\mathscr{D}}}.
\end{align}
Depending on the values of $\mathscr{C}$ and $\mathscr{D}$ (which also depend on $s$ in the general case), one is able to also optimise over $s$ to find the best order Trotter formula to use. For the remainder of this paper, however we set $s=2$ and these quantities reduce to
\begin{align}
\label{mainresult1}
n^* = \sqrt{\frac{\mathscr{C}}{\mathscr{D}}}
\end{align}
and
\begin{align}
\label{mainresult2}
D(n^*) = 2\sqrt{\mathscr{C}\mathscr{D}}.
\end{align}

As we shall show, $\mathscr{C}$ is fixed by the pair $\{H,t\}$, while $\mathscr{D}$ is determined by gate errors. Below we investigate some example noise models that influence $\mathscr{D}$.
\subsection{Mistimed control}
Trotterization fundamentally requires `switching'  between unitary gates: in our simplification this is thought of as switching each component Hamiltonian $H_j$ on for a specific  duration. Consider that the duration is increased (decreased) by a random number $\Delta_{ij}$, normally distributed around zero with variance $\sigma^2$. 
This makes 
\begin{align}
\mathcal{E}_{ij}^{\text{MTC}}(\rho)=e^{iH_j \Delta_{ij}}\rho e^{-iH_j \Delta_{ij}}.
\end{align}
Such imperfection is ubiquitous in the control of quantum systems, and an equivalent imperfection (laser intensity fluctuations) was cited as the dominant error source by Lanyon \emph{et al.} in a recent implementation of universal quantum simulation~\cite{LanyonHempelNigg2011}. Under this noise model, the total map $\mathcal{E}^{\text{faultyTrotter}}$, being a concatenation of random unitaries, is itself a random unitary map. Note that the magnitude of the errant time-shift $\Delta_{ij}$ is independent of $t$ or $n$, but the error map is more severe when $||H_i||$ is large, i.e. the simulator is driven `hard'. 
\begin{figure}[t!]
\includegraphics[width=\linewidth]{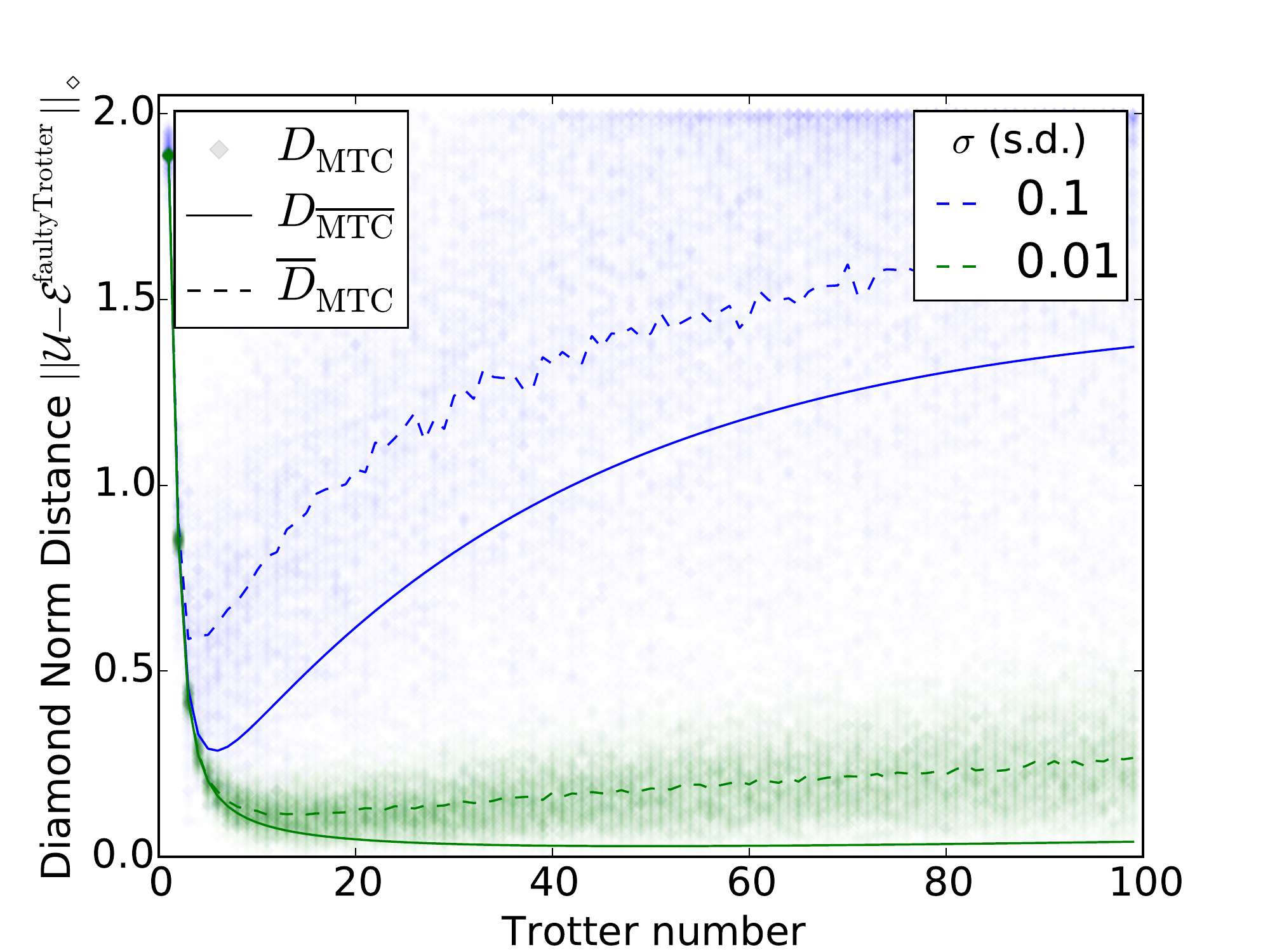}
\caption{(Color online) \color{black}Diamond norm distance versus Trotter number for a quantum simulation of the Hamiltonian $H =\sigma_x + \sigma_y$ over a duration $t=2$.  Each transparent marker represents a single Monte-Carlo simulated random unitary evolution of the simulator. The solid lines represent the performance of the averaged map and the dashed lines are the average of the individual Monte-Carlo runs -- lower is better. Note how the position of the averaged map is below the average position of the markers. The performance of the average map is better than the average performance of the maps of which it is composed. The two colours correspond to different noise characteristics of the mistimed control noise model. }
\label{advantage_of_averaging}
\end{figure}

As shown in Appendix~\ref{analytics}, by assuming $\norm{H_i[t/n+\Delta_{ij}]}\ll1$ one can find the difference between real and ideal supermatrices for this error model, finding
\begin{align}
\sqrt[n]{\mathbf{U}}-\mathbf{T}_{i}^{\text{fTss}}& \nonumber\\
=& \sum_{j<l}(\one\otimes [H_j,H_l]^*+[H_j,H_l]\otimes \one)\frac{t^2}{2n^2} \nonumber \\
&+i\sum_j(\one\otimes H^*_j-H_j\otimes \one)\Delta_{ij} \nonumber \\
&+\sum_{j<l}(\one\otimes H_j^*H_l^*+H_jH_l\otimes\one)\nonumber\\
&\times\left[\frac{t}{n}\Delta_{ij}+\frac{t}{n}\Delta_{il}+\Delta_{ij}\Delta_{il}\right]\nonumber \\
&+\frac{1}{2}\sum_{j}(\one\otimes H_j^{2^*}+H_j^2\otimes\one)\left[2\frac{t}{n}\Delta_{ij}+\Delta_{ij}^2\right]\nonumber \\
&-\sum_{jl}H_j\otimes H_l^*\left[\frac{t}{n}\Delta_{ij}+\frac{t}{n}\Delta_{il}+\Delta_{ij}\Delta_{il}\right]+\ldots
\end{align}
Note that the terms linear in $\Delta$ work to introduce `coherent' or reversible errors that may be averaged away because of our assumptions $\overline{\Delta_{ij}}=0$ and $\overline{\Delta_{ik}\Delta_{im}}=\delta_{km}\sigma^2$, hence
\begin{align}
\overline{\sqrt[n]{\mathbf{U}}-\mathbf{T}_{i}^{\text{fTss}}}=& \sum_{j<l}(\one\otimes [H_j,H_l]^*+[H_j,H_l]\otimes \one)\frac{t^2}{2n^2} \nonumber \\
&+\frac{1}{2}\sum_{j}(\one\otimes H_j^{2^*}+H_j^2\otimes\one)\sigma^2\nonumber \\
&-\sum_{jl}H_j\otimes H_l^*\sigma^2+\ldots
\label{averageddiff}
\end{align}
leaving the `incoherent' or irreversible errors (quadratic in $\Delta$) remaining. Clearly averaging can help reduce fluctuations introduced by the noise. This is expected, because of the convexity property of the norm. The resultant (less severe) map is denoted $\overline{\text{MTC}}$, ?for averaged mistimed control?. Figure~\ref{advantage_of_averaging} shows (through numerical simulations) how averaging can reduce the simulation error for this noise model, as measured by the diamond norm distance. Taking the norm of \eqref{averageddiff} gives the statistical distance, and employing the triangle inequality gives
\begin{align}
D^{ss}_{\overline{\text{MTC}}} \lessapprox \mathscr{A}\frac{t^2}{2n^2}+\mathscr{B}\sigma^2.
\end{align}
By the chaining property we reach
\begin{align}
D_{\overline{\text{MTC}}} \lessapprox \mathscr{A}\frac{t^2}{2n}+\mathscr{B}n\sigma^2,
\label{approximate_upper_bound}
\end{align}
with 
\begin{align}
\mathscr{A}=& \norm{\sum_{j<l}\left(\mathbb{I}\otimes[H_j,H_l]^*+[H_j,H_l]\otimes\mathbb{I}\right)}_? \nonumber \\
\mathscr{B}=&\norm{\frac{1}{2}\sum_j \left(\mathbb{I}\otimes H_j^{*^2}+H_j^2\otimes\mathbb{I}-2H_j\otimes H_j^*\right)}_?.
\end{align}
Comparing (26) with (17), we see that $\mathscr{A}t^2/2=\mathscr{C}$ and $\mathscr{B}\sigma^2=\mathscr{D}$. Note that we leave the choice of norm free here: $?\in \{\diamond,1,J,\ldots\}$. We made use of only the triangle inequality and the chaining property. As we will show, if $\mathscr{A}:\mathscr{B}$ happens to be invariant under choice of norm the optimum Trotter number is also an invariant, since it only depends on this ratio. Note also that these are norms of supermatrices: the supermatrices may need to be converted to another form in order that the norms are evaluated.

Applying the triangle inequality to $\mathscr{A}$ and $\mathscr{B}$ we note that the first quantity is upper bounded by the sum of at worst $\frac{1}{2}(k^2-k)$ norms, and the second by the sum of only $k$ norms. Since $k$ is polynomial in the particle number, the Trotter error (sometimes called the digital error) therefore grows only polynomially in the size of the simuland.

Taking $n\in\mathbb{R}$ for the moment, simple analysis yields i) the optimum Trotter number 
\begin{align}
n^*_{\overline{\text{MTC}}}=\sqrt{\mathscr{A}(2\mathscr{B})^{-1}}t\sigma^{-1},
\label{nstar}
\end{align}
ii) the statistical distance at this optimum 
\begin{align}
D_{\overline{\text{MTC}}}(n^*)=\sqrt{2\mathscr{A}\mathscr{B}}t\sigma
\label{Dnstar}
\end{align}
 and iii) the maximum simulation time after which the statistical distance is above the accepted level $D_{\text{max}}$, given by 
 \begin{align}
 t_{\text{max}}=D_{\text{max}}/(\sigma\sqrt{2\mathscr{A}\mathscr{B}}).
 \label{tmax}
 \end{align}
 The meaning of equations~\eqref{nstar},\eqref{Dnstar} and \eqref{tmax} is inherited from the choice of norm in the definition of $\mathscr{A}$ and $\mathscr{B}$. 
 
Of course in reality $n^*\in \mathbb{Z}$: the true optimum must be a whole number of Trotter steps, and so $n^*$ should be rounded in the direction of the sign of  $\lfloor n^* \rfloor\lceil n^* \rceil - \mathscr{A}t^2/(2\mathscr{B}\sigma^2)$. See Appendix~\ref{n_integer} for more details.

\begin{figure*}[t!]
\includegraphics[width=\textwidth]{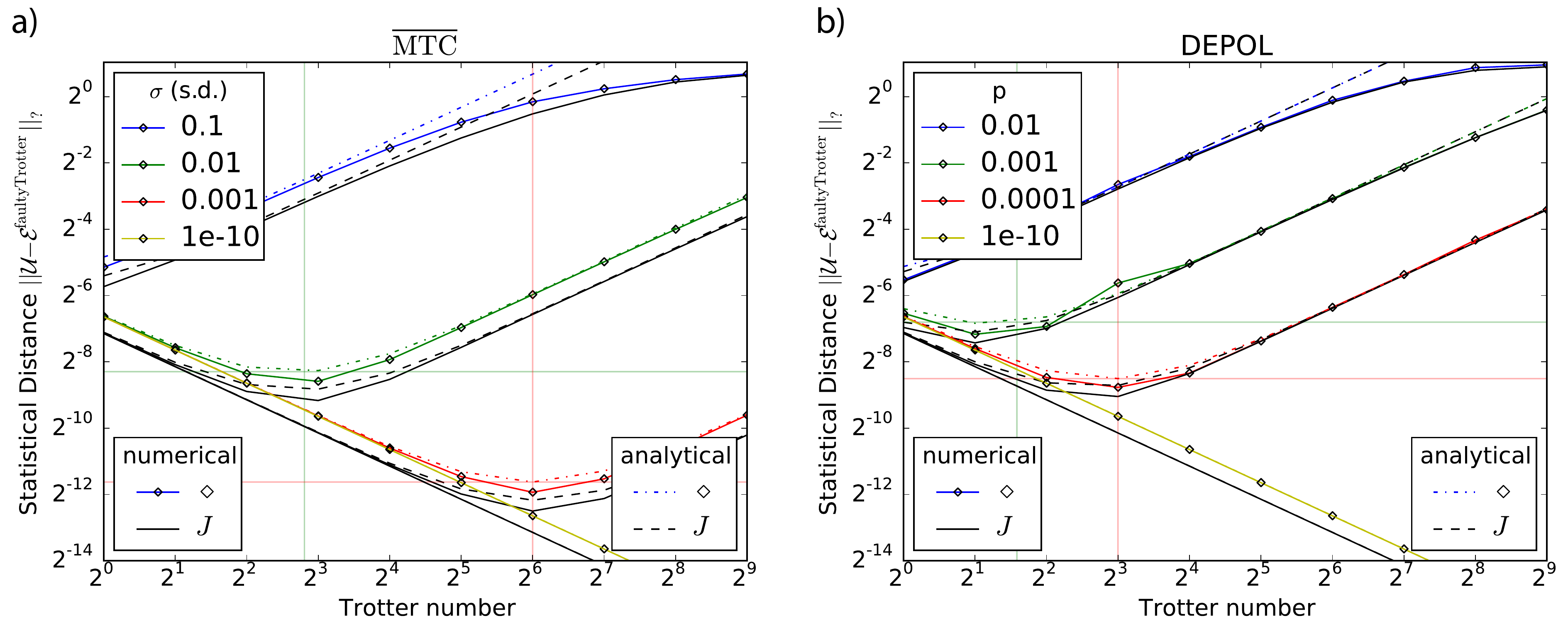}
\caption{\label{MTCfig}(Color online) {\color{black}Log-Log plot of the statistical distance versus the Trotter number for a faulty simulation of the Ising Hamiltonian ( Eq.~\eqref{IsingHamiltonian}), with $N=2$ and $t=0.1$.  The noise model is either a) timing errors (the $\overline{\text{MTC}}$ noise model) or b) depolarising noise (the DEPOL model). The horizontal axis represents the frequency of applied operations, and the vertical axis is related to the probability that the actual simulation could be distinguished from the ideal simulation (lower is better). Solid colored lines (numerically calculated) correspond to the diamond norm distance, and the dotted lines are approximate analytical formulae. The different colours correspond to either a) the standard deviation $\sigma$ of the noise or b) the probability $p$ of a depolarising error at each Trotter step, each shown in the legend. Black lines are the $J$-distance (solid lines are exact numerics, dotted lines are approximate analytics). The crosshairs show the analytically predicted optimum Trotter number and performance level at that optimum. Note the $n\rightarrow\infty$ behaviour shows the norms saturating to the completely noisy benchmark (see main text).}}
\label{loglog}
\end{figure*}

Although our formulae apply generally, as a case study we study a special case of the Ising Hamiltonian
\begin{align}
\label{IsingHamiltonian}
H_1 =\sum_r^N \sigma_z^r ,\quad H_2 = \sum_{\langle r,s\rangle} \sigma^r_x\sigma_x^{s}
\end{align}
with $\sigma^r_x$ and $\sigma^r_z$ being the Pauli matrices acting on qubit $r$ of $N$, with identity matrices implied on other qubits. The notation $\sum_{\langle r,s\rangle}$ denotes a sum over nearest neighbours. Note that $[H_1,H_2]\neq0$. As shown in Figure~\ref{MTCfig}a, our approximate analytical upper bound~\eqref{approximate_upper_bound} is a good fit for exact numerics in the correct parameter regime (for example we chose $t=0.1$ and $N=2$).

We note three regimes where the fit is worse: i) when $\norm{H_it/n}\approx 1$ (the perturbative expansion breaks down), ii) in the tradeoff region near the optimum Trotter number $n^*$ (here the chaining inequality is loose) and iii) when the distance measure approaches its maximum (the errors are saturating, and the chaining inequality between consecutive Trotter steps is loose). Despite the looseness of the fit in these areas, the location of $n^*$ and $D(n^*)$ are well captured. 

 Aside from combatting the gate noise via averaging, one can apply the time-energy freedom \eqref{tef}, noting that 
\begin{align}
a\neq 1 \Rightarrow D_{\overline{\text{MTC}}} \lessapprox \mathscr{A}\frac{t^2}{2n}+\mathscr{B}n(a\sigma)^2.
\end{align}
Clearly one can now take $a<1$, i.e. retard the simulator to decrease the absolute error at each Trotter step. This has the same effect as reducing $\sigma$-- the first term is unaffected.
\subsection{Trotter step induced depolarisation}
Next we study a different noise model, which we call the Trotter-step-induced depolarization, or DEPOL model. Depolarising noise is commonly used to phenomenologically model noise because it makes analytic results tractable. To gain an idea of how each operation might introduce generic noise into the simulator, we assume that after a single Trotter step the combined effect of imperfections in the applied gates is to depolarise the entire simulator uniformly. The faulty map is then
\begin{align}
\mathcal{E}^{\text{faultyTrotter}}(\rho)&=\bigcirc_i \mathcal{E}_i^{\text{DEPOL}}\circ(\ldots \mathcal{V}_2\circ\mathcal{V}_1)(\rho)\\
&=\bigcirc_i \mathcal{E}^{\text{fTss}}_i,
\label{stepper}
\end{align}
with 
\begin{align}
\mathcal{E}^{\text{DEPOL}}_{i}(\rho)=(1-p)\rho + \frac{p}{d}\mathbb{I}.
\end{align}
Here $\mathbb{I}$ is the $d\times d$ identity matrix. Note there is no dependence on any variable other than $p$, the probability of a depolarising error during one Trotter step. This quantity therefore captures the severity of the faulty control. Now
\begin{align}
\mathcal{E}^{\text{fTss}}-\sqrt[n]{\mathcal{U}}=&(1-p)\sqrt[n]{\mathcal{V}}\nonumber \\
&+p\E^{\rho\rightarrow\mathbb{I}/d}-\sqrt[n]{\mathcal{U}}\nonumber \\
=&(1-p)\left[\sqrt[n]{\mathcal{V}}-\sqrt[n]{\mathcal{U}}\right]\nonumber \\
&+p\left[\E^{\rho\rightarrow\mathbb{I}/d}-\sqrt[n]{\mathcal{U}}\right],
\end{align}
and we can use convexity to reach
\begin{align}
\norm{\mathcal{E}^{\text{fTss}}-\sqrt[n]{\mathcal{U}}}\leq &(1-p)\norm{\sqrt[n]{\mathcal{V}}-\sqrt[n]{\mathcal{U}}}\nonumber \\
&+p\norm{\E^{\rho\rightarrow\mathbb{I}/d}-\sqrt[n]{\mathcal{U}}}.
\end{align}
Now the first term follows trivially from the above calculation and equals $(1-p)\left.D^{\text{ss}}_{\overline{\text{MTC}}}\right|_{\sigma=0}$. The second term is also known (see above). This means that (again employing the chaining inequality over the $n$ iterations of the single step channel, and assuming $\norm{H_i}t/n\ll1$)
\begin{align}
D_{\text{DEPOL}}  \lessapprox (1-p)\mathscr{A}\frac{t^2}{2n}+np\left(2-\frac{2}{d^2}\right)
\end{align}
which we can simplify, in the case of $d\gg1$ to 
\begin{align}
D_{\text{DEPOL}}  \lessapprox (1-p)\mathscr{A}\frac{t^2}{2n}+2pn.
\label{statistical_distance_DEPOL}
\end{align}
Qualitatively this is very similar to the MTC model. One cannot however leverage the freedom in the time-energy correspondence to effectively reduce $p$, the characteristic of the noise, because \eqref{statistical_distance_DEPOL} is invariant under that transformation. One has
\begin{align}
n^*_{\text{DEPOL}}=t\sqrt{\frac{(1-p)\mathscr{A}}{2p(2-\frac{2}{d^2})}} 
\end{align}
or in the case of $d\gg1$ 
\begin{align}
n^*_{\text{DEPOL}}=t\sqrt{\frac{(1-p)\mathscr{A}}{4p}}.
\end{align}
Note that, as with above, one should round $n^*$ to the nearest integer in the direction of the sign of $\lfloor n^* \rfloor\lceil n^* \rceil - (1-p)\mathscr{A}t^2/4p$. One has 
\begin{align}
D_{\text{DEPOL}} (n^*)=2t\sqrt{\mathscr{A}p(1-p)}.
\end{align}
See Figure~\ref{MTCfig}b for comparison with numerical results, which show that the DEPOL noise model has qualitatively similar behaviour to the $\overline{\text{MTC}}$ noise model and also that our approximate analytical approach continues to provide a good fit for the true performance.

\subsection{Decoherence}
To model the overall noise of a simulator under perfect control, we depolarise the simulator in a manner that is independent of the number of operations. This means the number of Trotter steps does not affect the total noise characteristics. A very similar argument to the above gives
\begin{align}
\E^{\text{faultyTrotter}}-\mathcal{U}=& (1-p(t))[\mathcal{V}-\mathcal{U}]\nonumber \\
&+p(t)[\E^{\rho\rightarrow\mathbb{I}/d}-\mathcal{U}]
\end{align}
and now $p(t)$ increases over time. Typically $1-p(t)$ represents exponential decay of population and coherence with time~\cite{NielsenChuang2004}. The first term follows trivially and is equal to $(1-p)\left.D_{\overline{\text{MTC}}}\right|_{\sigma=0}$. We have
\begin{align}
D_{\text{DECOH}}\leq (1-p(t))\frac{\mathscr{A}t^2}{2n}+p(t)\left(2-\frac{2}{d^2}\right).
\end{align}
Once more employing the time energy freedom \eqref{tef} we find 
\begin{align}
D_{\text{DECOH}}\leq \left(1-p\left(\frac{t}{a}\right)\right)\frac{\mathscr{A}t^2}{2n}+p\left(\frac{t}{a}\right)\left(2-\frac{2}{d^2}\right);
\end{align}
note that accelerating the simulator $a>1$ will improve performance if $p(t)$ is monotonically increasing. Note here there is no optimum Trotter number, because the noise is not worsened by increasing the number of operations.
\section{Discussion}
By modelling some simple imperfections in the control of a Trotterized universal quantum simulator, we have shown how the accuracy of the quantum channel that is applied depends on various parameters. These parameters include those outside of direct experimental control, such as the severity of environmental decoherence or the spread in control timing; and also controllable quantities such as the Trotter number and the simulator/simuland time ratio. We suggested ways in which these quantities can be optimised over in order to improve the accuracy of the simulation, which we quantified by calculating the statistical distance to the ideal map. The optimisation implies only that the uncontrollable quantities be estimated from control experiments, and that $\mathscr{C}$ (simply related to $\mathscr{A}$ which measures the total non-commutativity of the local pieces of the Hamiltonian) and $\mathscr{D}$ (which measures the control error per Trotter step) be calculated only once (possibly numerically). We discussed the choice of a number of norms for this purpose, including the worst-case induced-trace norm and stabilised (diamond) norm, as well as the average-case $J$-distance.

In particular we found there to be a finite optimum Trotter number to employ, setting a maximum performance and a maximum simulation time. By way of a general argument, we have shown these features to be generic to faulty simulators operated with a Trotterized (or similar) algorithm. In order to predict these quantities, one can appeal to a microscopic model of the simulator -- having a sufficiently good model for a physical system is arguably a prerequisite for using it as a computational device~\cite{HorsmanStepneyWagner2014}. 

{\color{black}The problem of noise in quantum computers has, in one sense, been solved by the error-correction threshold theorems~\cite{AharonovBen-Or1997,Kitaev1997,AliferisGottesmanPreskill2006,KnillLaflammeZurek1998}. An error correction threshold is a critical value of a measure of the accuracy of quantum control. Once it has been surpassed, error correction techniques work to decrease the overall net error. These theorems show that, when experimental operations become clean enough, encoding logical qubits in a larger number of redundant physical qubits allows the error in the overall computation to be suppressed at will~\cite{Terhal2013}.  But as interest in simulators grows~\cite{LanyonWhitfieldGillett2010}, and small scale (say 64 qubit) devices begin to appear, the noise problem remains until many-thousand-qubit devices with sub-threshold physical error rates can be engineered. Even after this is achieved, imperfections will not have completely disappeared: the question of how the accuracy of overall computation depends on the necessarily finite \emph{residual} (i.e. error-corrected) error rate~\cite{Raussendorf2012} remains of high importance, and our analyses here will still apply.}

We will briefly comment on two recent experimental demonstrations in superconducting systems. The Martinis group perform an investigation of fermionic models with four transmon qubits~\cite{BarendsLamataKelly2015}: but they fail to control the error in the simulation at all, operating in the gate-error-dominated regime $n > n^*$. The Wallraff group, investigating interacting spin systems using only two transmon qubits, successfully sweep the Trotter number through the optimum point~\cite{SalatheMondalOppliger2015}, for various times $t$. By increasing the total simulation time they effectively increase the Trotter error until it competes with the other errors. Our results predict that, if both the Hamiltonian and error per Trotter step (i.e. $\mathscr{C}$ and $\mathscr{D}$) are kept constant then $n^*/t$ should be a constant: the experimental results are in agreement with this prediction. The worst trace distance in these experiments was of order $10^{-1}$.

These experiments show the difficulty in achieving good confidence in quantum simulators, especially as the number of qubits is increased. Using our results ~\eqref{mainresult1} and~\eqref{mainresult2} we estimate that for the Wallraff experiment (fixing the Hamiltonian and simulation time) reaching the sorts of precision we are accustomed to with modern classical computers (the current standard `single precision' is $2^{-24}\approx 10^{-7}$~\cite{Overton2001}) will require a factor of at least $10^6$ more Trotter steps and therefore a $10^{12}$ improvement in the error per Trotter step. This is due to the fact that the overall error is proportional to the square root of the error per Trotter step. 

Furthermore as pointed out in Ref.~\cite{ClarkMetodiGasster2009}, a full simulation using readout with the phase estimation algorithm calls for simulation not only of $U$ but of $U^{2m}$, further increasing the required number of Trotter steps and therefore precision per step. As Wecker \emph{et al.} show~\cite{WeckerBauerClark2014}, the number of logical gates per Trotter step (for example in a quantum chemistry simulation) is roughy $O(N^5)$. This fact means that the error per gate may need to be several orders of magnitude smaller still. 

That simulations are only likely to provide novel insight when featuring many dozens of qubits implies that yet another improvement in gate quality is necessary. Consider an $N=64$ qubit simulation with nearest neighbour interactions. Compared to the recent $N=2$ experiment~\cite{SalatheMondalOppliger2015}, there will be a thousandfold increase in the number of commutators between the $H_i$; many of these may be expected to vanish but nevertheless we might expect $\mathscr{A}$ (and therefore $\mathscr{C}$) to increase by two or even three orders of magnitude, requiring an improvement in $\mathscr{D}$ of comparable magnitude to compensate.

Future work will investigate the multiparameter optimisation when all noise types are present, the use of alternative Trotter-type approximants, and the performance of simulators when the set of interesting states and applied measurements is chosen from a restricted set.

\begin{acknowledgments}
We thank Yuichiro Matsuzaki and Kae Nemoto for helpful discussions, and Viv Kendon for comments on an earlier version of this manuscript.

\appendix

{\color{black}
\section{Analytics\label{analytics}}
}
We are interested in the difference between the faulty Trotter map described by the $\overline{\text{MTC}}$ noise model and the ideal map. It is convenient here to choose the super-matrix representation. For a single shot map over a single Trotter step 
\begin{widetext}
\begin{align}
\sqrt[n]{\mathbf{U}}-\mathbf{T}_{i}^{\text{fTss}} &=& e^{i\sum_j H_jt/n}\otimes e^{-i\sum_j H^*_jt/n}- {\color{black} \prod_j e^{iH_j(t+n\Delta_{ij})/n}\otimes \prod_j e^{-iH_j^*(t+n\Delta_{ij})/n}}\nonumber \\
&=&\left(\one+\frac{it}{n}\sum_jH_j-\frac{t^2}{2n^2}\left\{\sum_{j>l}+\sum_{j<l}+\sum_{j=l}\right\}H_jH_l+\ldots\right)\nonumber \\
&&\otimes \left(\one-\frac{it}{n}\sum_jH^*_j-\frac{t^2}{2n^2}\left\{\sum_{j>l}+\sum_{j<l}+\sum_{j=l}\right\}H_j^*H_l^*+\ldots\right) \nonumber \\
&&-{\color{black} \left(\one+i\sum_jH_j\left[\frac{t}{n}+\Delta_{ij}\right]-\frac{1}{2}\left\{\sum_{j=l}+2\sum_{j<l}\right\}H_jH_l\left[\frac{t}{n}+\Delta_{ij}\right]\left[\frac{t}{n}+\Delta_{il}\right]+\ldots\right)}\nonumber \\
&&{\color{black} \otimes \left(\one-i\sum_jH^*_j\left[\frac{t}{n}+\Delta_{ij}\right]-\frac{1}{2}\left\{\sum_{j=l}+2\sum_{j<l}\right\}H_j^*H_l^*\left[\frac{t}{n}+\Delta_{ij}\right]\left[\frac{t}{n}+\Delta_{il}\right]+\ldots\right)}\\
&=& - \frac{t^2}{2n^2}\left\{\sum_{j>l}+\sum_{j<l}+\sum_{j=l}\right\}\left(\one\otimes H_j^*H_l^* + H_jH_l\otimes\one\right) {\color{black} -i\sum_j\Delta_{ij}( H_j\otimes\one-\one\otimes H_j^*)}\nonumber \\
&&{\color{black}+\frac{1}{2}\left\{\sum_{j=l}+2\sum_{j<l}\right\}(\one\otimes H_j^*H_l^*+H_jH_l\otimes\one)\left[\frac{t^2}{n^2}+\frac{t}{n}\Delta_{ij}+\frac{t}{n}\Delta_{il}+\Delta_{ij}\Delta_{il}\right]}\nonumber \\
&&{\color{black}-\sum_{jl}H_j\otimes H_l^*\left[\frac{t}{n}\Delta_{ij}+\frac{t}{n}\Delta_{il}+\Delta_{ij}\Delta_{il}\right]}+\ldots
\end{align}
which, after employing the identity 
$$-\left\{\sum_{j>l}+\sum_{j<l}+\sum_{j=l}\right\}+\left\{\sum_{j=l}+2\sum_{j<l}\right\}H_jH_l\equiv\left\{\sum_{j<l}-\sum_{j>l}\right\}H_jH_l\equiv\sum_{j<l}[H_j,H_l],$$
gives
\begin{align}
\sqrt[n]{\mathbf{U}}-\mathbf{T}_{i}^{\text{fTss}} =& \sum_{j<l}(\one\otimes [H_j,H_l]^*+[H_j,H_l]\otimes \one)\frac{t^2}{2n^2}+i\sum_j(\one\otimes H^*_j-H_j\otimes \one)\Delta_{ij} \nonumber \\
&+\sum_{j<l}(\one\otimes H_j^*H_l^*+H_jH_l\otimes\one)\left[\frac{t}{n}\Delta_{ij}+\frac{t}{n}\Delta_{il}+\Delta_{ij}\Delta_{il}\right]\nonumber \\
&+\frac{1}{2}\sum_{j}(\one\otimes H_j^{2^*}+H_j^2\otimes\one)\left[2\frac{t}{n}\Delta_{ij}+\Delta_{ij}^2\right]\nonumber \\
&-\sum_{jl}H_j\otimes H_l^*\left[\frac{t}{n}\Delta_{ij}+\frac{t}{n}\Delta_{il}+\Delta_{ij}\Delta_{il}\right]+\ldots
\end{align}
which is the result in the main text.
\end{widetext}

{\color{black}
\section{Numerics}
}
The diamond norm distance between two unitary maps described by matrices $U$ and $V$ is simply the diameter of the smallest enclosing circle of the eigenvalues of $UV^\dagger$~\cite{JohnstonKribsPaulsen2009}.  In our numerical work we used {\sc python} code from Minase~\cite{smallestenclosingcircle}. 

To calculate the diamond norm for a general map expressed in supermatrix, Choi or Jamiolkowski form, we used a semidefinite programming~\cite{GoemansWilliamson2004,BoydVandenberghe2009} algorithm due to Watrous~\cite{Watrous2012,Watrous2009}, implemented by Johnston in a {\sc matlab} package called {\sc qetlab}~\cite{qetlab}.

\subsection{Total noise and depolarisation}
What is the supermatrix form of $\mathcal{E}^{\rho\rightarrow\mathbb{I}/d}(\rho)=\one/d$? One can exploit the trace of the density matrix to get the following supermatrix:
\begin{align}
\mathbf{T}^{\rho\rightarrow\mathbb{I}/d}= 
\left(
\begin{array}{ccccccc}
\frac{1}{d} & 0 & \ldots & 0 & \frac{1}{d}&0&\ldots\\
0 & 0 & \ldots & 0 & 0&0&\ldots\\
\vdots&\vdots&\ddots&\vdots&\vdots&\vdots&\ddots \\
0 & 0 & \ldots & 0 & 0&0&\ldots \\
\frac{1}{d} & 0 & \ldots & 0 & \frac{1}{d}&0&\ldots \\
0 & 0 & \ldots & 0 & 0&0&\ldots\\
\vdots&\vdots&\ddots&\vdots&\vdots&\vdots&\ddots \\
\end{array}
\right)
\end{align}
each ellipsis denotes that there are $d$ zero entries between the nonzero elements. Clearly each off-diagonal is annihilated and each diagonal goes to $(\text{Tr}\rho)/d$. 
From here we can easily construct a supermatrix representation of $\mathcal{E}^{\text{DEPOL}}$, for example by scaling $\mathbf{T}^{\rho\rightarrow\mathbb{I}/d}$ by $p$ and adding $(1-p)\one_{d^2\times d^2}$.

\subsection{Averaged Mistimed Control}
The quantum channel under consideration is what we call the \emph{averaged} mistimed control ($\overline{\text{MTC}}$) model. It is the average map that is applied when the timing suffers a nonzero spread.
\begin{align}
\overline{\mathbf{T}}^{\text{faultyTrotter}} = \prod_i \overline{\mathbf{T}}_i^{\text{nTss}}=\prod_i\prod_j \overline{\mathbf{T}}_{\mathcal{E}_{ij}}\mathbf{T}_{\mathcal{U}_{ij}};
\end{align}
here we have exploited the independence of the random (matrix) variables to distribute the expectation over the concatenation of channels. Now for the $\overline{\text{MTC}}$ model the interleaved maps can be written 
\begin{align}
\overline{\mathbf{T}}_{\mathcal{E}_{ij}}: \rho\rightarrow & \int  p(\Delta_{ij}) \e{i\hat{H}_j\Delta_{ij}}\rho\e{-i\hat{H}_j\Delta_{ij}}d\Delta_{ij}
\end{align}
with the probability distribution $p(\Delta_{ij})$ taken as a product of independent Gaussians with zero mean and standard deviation of $\sigma$.  Now, expanding the density matrix $\rho$ in the energy eigenbasis, each operation merely generates a phase and we can evaluate the integral (here we suppress the Hamiltonian index $j$ for clarity):
\begin{widetext}
\begin{align}
\rho\rightarrow&\frac{1}{\sqrt{2\pi}\sigma}\int \e{-\Delta_{ij}^2/\sigma^2}\e{i\hat{H}\Delta_{ij}}\left(\sum_{mn} \tilde{\rho}_{mn}|E_m\rangle\langle E_n|\right)\e{-i\hat{H}\Delta_{ij}}d\Delta_{ij}\\
=&\frac{1}{\sqrt{2\pi}\sigma}\int \e{-\Delta_{ij}^2/\sigma^2}\e{iE_m\Delta_{ij}}\e{-iE_n\Delta_{ij}}\left(\sum_{mn} \tilde{\rho}_{mn}|E_m\rangle\langle E_n|\right)d\Delta_{ij}\\
=&\sum_{mn} \e{-\frac{1}{2}(E_m-E_n)^2\sigma^2}\tilde{\rho}_{mn}|E_m\rangle\langle E_n|.
\end{align}
\end{widetext}
We defined $\tilde{\rho}_{mn}=\bra{E_m}\rho\ket{E_n}$ with $H\ket{E_m}=E_m\ket{E_m}$. Observe how the map acts on energy eigenstates: as an element-wise product $\tilde{\rho}\rightarrow \Lambda \cdot \tilde{\rho}$ with a matrix having entries $\Lambda_{mn}=\e{-\frac{1}{2}(E_m-E_n)^2\sigma^2}$. If we define $W$ as the unitary matrix changing from the canonical into the energy eigenbasis, then the map acts on any $\rho$ as $\rho\rightarrow W^\dagger(\Lambda\cdot W\rho W^\dagger)W$. 

The supermatrix $\mathbf{T}_{\Lambda}$ corresponding to the elementwise product can be found by reshaping $\Lambda$ into a vector and then constructing a diagonal matrix from this vector. The advantages of using a supermatrix is the composition of channels is simply matrix multiplication. The map can be defined for any density matrix in the following way
\begin{align}
\overline{\mathbf{T}}_{\mathcal{E}_{ij}}\vec{\rho}= \mathbf{T}_{W^\dagger} \mathbf{T}_{\Lambda} \mathbf{T}_{W} \vec{\rho}
\end{align}
and $\mathbf{T}_{W} = W\otimes W^*$ and so on.

By using these techniques, one is able to build up the total (averaged) faulty Trotter map in supermatrix form, which is convenient for numerical calculations.

{\color{black}
\section{Finding $n^*$ as an Integer\label{n_integer}}
}
Using calculus, i.e. assuming $n$ to be a real number, will generally give a non integer solution for $n^*$. We can always round this up (or down) to get an integer, and in the worst case we may round in the wrong direction. Here we briefly show how to make sure the integer chosen is the correct one. To ensure the lowest value for $D$, clearly we should choose the `rounded down' value $\lfloor n^* \rfloor$ or the `rounded up' value $\lceil n^* \rceil$ when the quantity
\begin{align}
D(\lceil n^* \rceil )- D(\lfloor n^* \rfloor) 
\end{align}
is positive and negative respectively. When it is exactly zero, we can choose either. Let $\lfloor n^* \rfloor=k$ and let 
\begin{align}
D(k) = \frac{\mathscr{C}}{k}+\mathscr{D}k.
\end{align}
and our condition simplifies to 
\begin{align}
k(k+1)-\frac{\mathscr{C}}{\mathscr{D}}.
\end{align}
This justifies the claim in the main text that one should round $n$ in the direction of the sign of $\lceil n^* \rceil \lfloor n^* \rfloor - (\mathscr{C}/\mathscr{D})$.

\end{acknowledgments}

\bibliography{/Users/georgeknee/Documents/paper_library/gck_full_bibliography}

\end{document}